\documentclass[useAMS,usegraphicx,usenatbib]{mn2e}
\usepackage{aas_macros}     
\usepackage{amssymb}        

\usepackage{fixltx2e}[1999/12/01]
\usepackage{rotating}

\newcommand{\Mhost}{ \ensuremath{M_{\rm host}} }
\newcommand{\rhost}{ \ensuremath{R_{\rm host}} }
\newcommand{\fmass}{ \ensuremath{M_{\rm sub}/M_{\rm host}} }
\newcommand{\Gpc}{ \ensuremath{\rm Gpc} }
\newcommand{\Mpc}{ \ensuremath{\rm Mpc} }
\newcommand{\kms}{ \ensuremath{\rm kms} }
\newcommand{\kpc}{ \ensuremath{\rm kpc} }
\newcommand{\Mass}{ \ensuremath{ h^{-1} M_{\odot}} }

\newcommand{\gsim}{\gtrsim} 

\title[The Fate of Substructures in Dark Matter Haloes]
{The Fate of Substructures in Cold Dark Matter Haloes}
\author[Angulo  et al.]{
\parbox[h]{160mm}{
R. E. Angulo\thanks{\rm E-mail: raul.angulo@durham.ac.uk},
C. G. Lacey,
C. M. Baugh,
C. S. Frenk.
}\vspace{6pt}\\
Institute for Computational Cosmology, Department of Physics, 
University of Durham, South Road, Durham, DH1 3LE, UK. 
\vspace*{-0.5cm}}

\begin{document}
\date{\today}
\pagerange{\pageref{firstpage}--\pageref{lastpage}} \pubyear{2008}
\maketitle
\label{firstpage}

\begin{abstract}
We use the Millennium Simulation, a large, high resolution N-body
simulation of the evolution of structure in a $\Lambda$CDM cosmology,
to study the properties and fate of substructures within a large
sample of dark matter haloes.  We find that the subhalo mass function
departs significantly from a power law at the high mass end. We also
find that the radial and angular distributions of substructures depend
on subhalo mass. In particular, high mass subhaloes tend to be less
radially concentrated and to have angular distributions closer to the
direction perpendicular to the spin of the host halo than their less
massive counterparts. We find that mergers between subhaloes
occur. These tend to be between substructures that were already
dynamically associated before accretion into the main halo. 
For subhaloes larger than $0.001$ times the mass of the host halo, 
it is more likely that the subhalo will merge with the central or main
subhalo than with another subhalo larger than itself. For lower masses, 
subhalo-subhalo mergers become equally likely to mergers with the main 
subhalo. Our results have implications for the
variation of galaxy properties with environment and for the treatment
of mergers in galaxy formation models.
\end{abstract}


\section{Introduction}

The presence of substructures within dark matter haloes is a distinctive
signature of a universe where structures grow hierarchically. Low mass objects
collapse at high redshift, and then increase their mass by smooth accretion of
dark matter or by merging with other haloes.  Once a halo is accreted by a
larger one, its diffuse outer layers are rapidly stripped off by tidal forces.
However, the core, which is much denser, generally survives the accretion event
and can still be recognized as a self bound structure or subhalo within the
host halo for some period of time afterwards.

In early N-body simulations, haloes appeared as fairly smooth objects
\citep{Frenk85,Frenk88}. However, as the attainable mass and force resolution
has increased, subhaloes have been identified and their properties studied in
detail by many authors over the past decade
\citep[e.g.][]{Ghigna1998,Ghigna2000, Tormen1998, Moore1999,Klypin1999a,
Klypin1999b,Springel2001, Stoehr2002,
deLucia2004,Gao2004,Nagai2005,Shaw2007,Diemand08,Springel08}. The properties of the subhalo
population have important implications for galaxy formation, dark matter
detection experiments and weak lensing. For instance, subhaloes are expected to
host satellite galaxies within groups and clusters and their evolution once
inside the host could give rise to observable changes. In particular, a merger
between two substructures could trigger an episode of star formation or a
morphological transformation \citep[e.g.][]{Somerville1999}.

In spite of this, the merger history of subhaloes remains relatively
unexplored. This is a challenging problem which demands a simulation with high
mass and force resolution. In particular, obtaining a statistical sample of
mergers involving the largest substructures requires a large sample of host
haloes. Most studies of substructure in halos have focused on resimulating, at
very high resolution, a small number of halos selected from a larger, lower
resolution simulation.  However, by studying only a few haloes, important
aspects related to variations produced by differences in the accretion and
merger histories of haloes, as well as any influence of the environment, could
remain hidden. This approach may also introduce systematic biases arising from
the criteria used to select the haloes to be resimulated.

In this paper, we overcome these problems by using the largest dark matter
simulation published to date, the Millennium Simulation (MS,
\citealt{Springel2005a}).  The MS provides a large cosmological sample of dark
matter haloes and associated substructures spanning a considerable range in
mass, allowing us to assess robustly the properties and fate of the subhalo
population. We complement our results with a higher resolution simulation of a
smaller volume (hereafter HS) which has a particle mass almost ten times
smaller than that used in the MS (Jenkins~et.al, in prep).

The layout of this paper is as follows. In Section
\ref{sub:sec:method}, we briefly describe the simulations used in this
work along with the properties of our halo and subhalo catalogues. In
Section \ref{sub:sec:prop} we investigate some general properties of
subhaloes, namely their mass function, radial distribution and spatial
orientation with respect to their host halo. The exploration of
substructure mergers and destruction is presented in Section
\ref{sub:sec:merger}. Finally, we summarize our findings in Section
\ref{sub:sec:conc}.

\section{Method} \label{sub:sec:method}

In this section we describe the N-body simulations we have  analyzed in
this work. We also discuss the identification and characterization of the halo
and subhalo catalogues.

\subsection{N-body Simulations} \label{sub:sec:method:mill}

The main simulation on which our analysis is based is the Millennium Simulation
\citep{Springel2005a}. The MS covers a comoving volume of $0.125 \,h^{-3}
\Gpc^{3}$ of a $\Lambda$CDM universe in which the dark matter component is
represented by $2160^3$ particles. The assumed cosmological parameters are in
broad agreement with those derived from joint analyses of the 2dFGRS galaxy
clustering \citep{Percival2001} and WMAP1 microwave background data
\citep{Spergel2003, Sanchez2006}, as well as with those derived from WMAP5 data
\citep{Komatsu2008}. In particular, the total mass-energy density, in units of the
critical density, is $\Omega_{\rm m} = \Omega_{\rm dm}+\Omega_{\rm b}=0.25$,
where the two terms refer to dark matter and baryons, with $\Omega_{\rm
b}=0.045$; the amplitude of the linear density fluctuations in $8 h^{-1} \Mpc$
spheres is $\sigma_8=0.9$; and the Hubble constant is set to $h = H_0/(100
\,\kms^{-1} \Mpc^{-1}) = 0.73$. The particle mass is $m_p = 8.6 \times
10^{8}\,\Mass$ and the Plummer-equivalent softening of the gravitational force
is $\epsilon = 5\, h^{-1} \kpc$.

To complement our results and to test for numerical effects we have also
employed another simulation with better mass resolution to which we refer as
HS. This simulation follows $900^3$ dark matter particles in a $\Lambda$CDM
cube of side $100\,h^{-1}\Mpc$.  The HS assumes the same cosmological
parameters as the MS. However, the smaller box yields a smaller particle mass,
$m_p = 9.5 \times 10^{7}\,\Mass$, so objects of a given mass are resolved with
almost 10 times more particles than in the MS.  Finally, in the HS the
softening length is $\epsilon = 2.4\,h^{-1} \kpc$.
  
The MS and HS were carried out using a memory efficient version of the {\tt
Gadget-2} code \citep{Springel2005b}.

\subsection{Halo and Subhalo catalogues} \label{sub:sec:method:haloes}

In both simulations, particle positions and velocities are written at 64 output
times which, for $z < 2$, are roughly equally spaced in time by $300$ Myr. In
each of these outputs we have identified dark matter haloes using the
friends-of-friends (FoF) algorithm \citep{Davis1985}, with a linking length of
0.2 times the mean interparticle separation. The volume and particle number of
the MS provide a unique resource of well resolved haloes to study. By way of
illustration, there are $90891$ haloes at $z=0$ with mass in excess of
$5.4\times10^{12}\,\Mass$ (one of the bins we use below), which corresponds to
$6272$ particles; at $z=1$ the number of haloes in excess of this mass is still
$61481$. On the cluster-mass scale, for example, there are $356$ haloes at
$z=0$ which are more massive than $4\times10^{14}\,\Mass$, corresponding to
$464576$ particles.

Well resolved FoF haloes are not smooth, but contain a considerable amount of
mass in the form of substructures. These dark matter clumps or ``subhaloes" are
identified and catalogued using a modified version of the subhalo finder
algorithm, {\tt SUBFIND}, originally presented in \cite{Springel2001}. The
statistics of the subhalo catalogue are impressive. At $z=0$ {\tt SUBFIND}
lists $339840$ structures with more than 200 particles in the MS within haloes
of at least $5.4\times10^{12}\,\Mass$. At $z=1$ there are $194629$
substructures with the same characteristics.  Note that {\tt SUBFIND} not only
finds substructures within a FoF halo, but it is also capable of identifying
substructures within substructures.

An important issue for studies of substructures is the definition of the
boundary and position of the host halo. In our analysis, the centre of the halo
is defined as the position of the most bound particle (i.e. usually the one
possessing the minimum gravitational potential). This choice for the halo
position agrees, within the softening length, with that found by a shrinking
sphere algorithm \citep{Power2003} for 93\% of the haloes that are resolved
with 450 or more particles. As shown by \cite{Neto2007}, the 7\% of cases in
which the two methods disagree are due to the FoF algorithm artificially
linking multiple structures. In these cases the position of the most bound
particle provides a more robust identification of the centre, as noted by
\cite{Neto2007}.

We define the halo boundary as the sphere of radius $r_{200}$ which contains a
mean density of 200 times the critical density, $\rho_{\rm crit}$. Therefore,
the mass of the halo is:

\begin{equation}
M_{200} = \frac{4}{3} \pi 200 \rho_{\rm crit} r_{200}^3 .
\end{equation}      

We keep in our catalogues only subhaloes within this sphere.  Although the
choice of the factor of 200 is motivated by the spherical collapse model in a
Einstein-de-Sitter universe, it is somewhat arbitrary for our $\Lambda$CDM
simulations. However, the $r_{200}$ definition has the advantage of being
independent of both redshift and cosmology.  Moreover, it has became a de facto
standard in studies of substructures. Nevertheless, we have tested our results
against other definitions of the halo boundary without finding any qualitative
differences. In the following, when we refer to the mass and radius of a host
halo, we always mean $M_{200}$ and $r_{200}$. 

Finally, we build merger trees using an algorithm similar to that described by
\citet{Springel2005a} which follows the evolution of subhaloes. In this way, we
can identify the haloes and subhaloes that will be involved in a merger during
a subsequent snapshot. Note that these merger trees are constructed using only
the information contained in the FoF and {\tt SUBFIND} catalogues, and there is
no attempt to force mass conservation, as would be required if the merger trees
were to be used in a galaxy formation code \citep[see][]{Harker2006}. The
descendant of a subhalo is defined as the structure that contains the majority
of the 10 percent most bound particles from the subhalo.  When two satellite
subhaloes have the same descendant in a following snapshot, we tag such an
event as a substructure merger.

\begin{figure*} 
\begin{center}
\includegraphics[width=17cm]{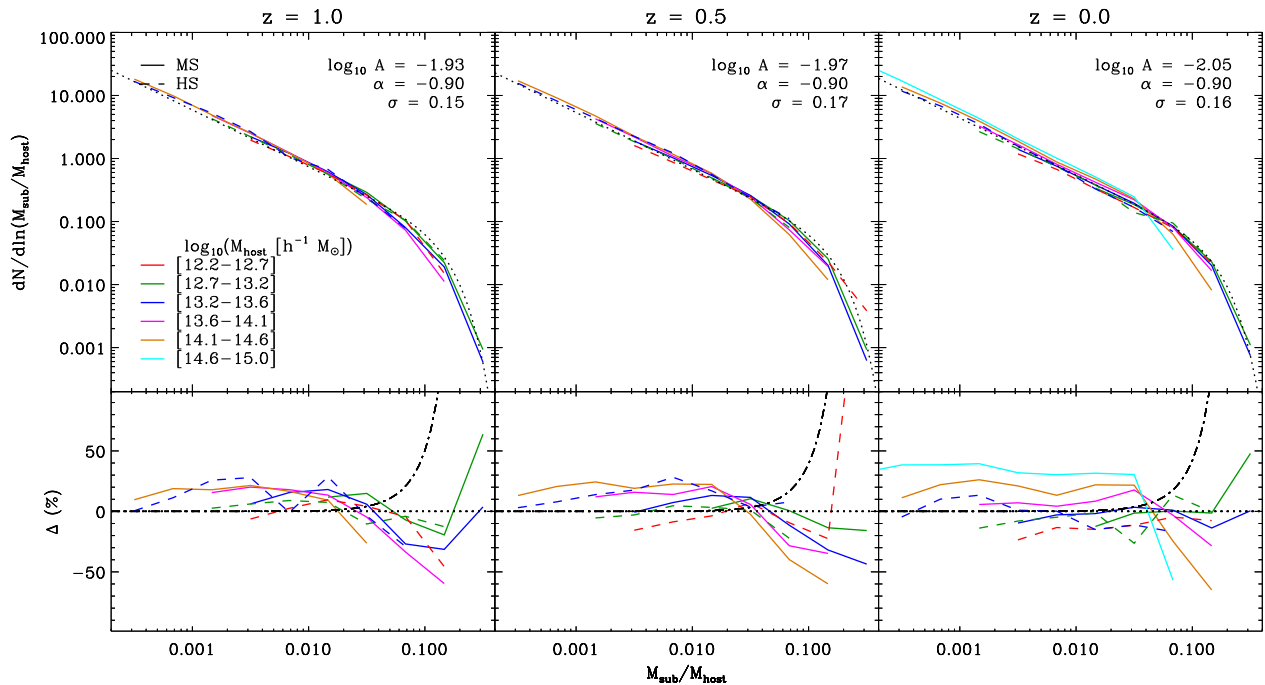}
\caption[Differential number of substructures per host halo as a
function of their mass relative to that of the host halo.]{ 
{\it Top row}: Differential number of substructures per host halo as a
function of their mass relative to that of the host halo,
$\fmass$. Note this is the mass of the subhalo at the redshift
labelled, in some cases after substantial stripping of mass has taken
place. Solid lines show the results from the MS while dashed lines
show the results from the HS.  In both cases, lines of different
colours show the subhalo mass function in host haloes of different
masses (as indicated by the legend). Each column shows a different
redshift, as labelled. At each redshift, the dotted lines display the
overall best fit of our model, Eq.~\ref{sub:eq:submf}, with the parameter
values given in the legend.  Parameters of the fits to individual mass
bins at $z=0$ are listed in Table 1.  {\it Bottom row}: Relative difference
between the overall best fit and measurements of the subhalo mass
function for different host masses. The dot-dashed line shows the
difference between our model, Eq.~\ref{sub:eq:submf}, and a power-law
fit. Only results for subhaloes which are resolved with more than $50$
particles are shown.
\label{sub:fig:submf}}
\end{center}
\end{figure*}

\section{SubHalo properties} \label{sub:sec:prop}

Before presenting our results regarding subhalo mergers, we consider 
some general properties of the subhalo population. Although some of
these properties  have been studied by previous authors, the large volume 
and high resolution of the MS and HS reveal some features which were 
inaccessible to earlier work.  Furthermore, the knowledge of the subhalo 
properties will help us to understand the results presented in the 
next section.

\subsection{Subhalo mass function}

We first consider the distribution of subhalo masses - the subhalo
mass function.  The top panels of Fig.~\ref{sub:fig:submf} show the mean
number of substructures within dark matter haloes, per host halo, per
logarithmic interval in subhalo mass. The results are displayed as a
function of subhalo mass relative to the mass of the halo in which it
resides, $\fmass$. In this way we can easily compare results across a
range of halo masses. In the ranges of overlap, the results from the
MS and HS agree well; this provides a useful, but limited, test of
the convergence of our results.

For the redshifts plotted in Fig.~\ref{sub:fig:submf} there is only a
small variation of the subhalo mass function with host halo
mass. Indeed, a universal function describes the behaviour reasonably
well over the range of subhalo mass resolved by our simulations:

\begin{equation}
\frac{\rm dN}{\rm d\ln(M_{\rm sub}/\Mhost)} = A\, 
\left( \frac{ M_{\rm sub}} {\Mhost} \right)^{\alpha} 
\, \exp \left[ - \frac{1}{\sigma^2} 
\left(  \frac{ M_{\rm sub}} {\Mhost} \right)^2 \right], 
\label{sub:eq:submf}
\end{equation}

\noindent where $N$ is the number of subhaloes per host halo. 
The values of $A$, $\alpha$ and $\sigma$ in this overall fit at 
each redshift are
given in the legend of Fig.~\ref{sub:fig:submf}. For this overall fit, we
have forced the slope $\alpha$ to have the same value independently of
redshift. In general, we find that $\alpha=-0.9$ is a good
approximation to the best fit from $z=0$ to $z=2.5$. It is also
important to note that the power-law fit widely used in the
literature, (e.g.  \citealt{Gao2004}) is only valid over a limited
range of fractional subhalo masses, $\fmass < 0.04$. We also see that
the maximum subhalo mass for which a power-law successfully describes
the mass function decreases at higher redshifts, $\fmass \sim
0.015$ at $z=1$ and $\fmass \sim 0.04$ at $z=0$.  The bottom panels of
Fig.~\ref{sub:fig:submf} show the relative difference between the fit
given by Eq.~(\ref{sub:eq:submf}) and the mass function of subhaloes
measured in host haloes of different masses.

\begin{table}
\begin{center}
\begin{tabular}{cccccc}
\hline
& $M_{\rm host}$           & $\log_{10} A$ & $\alpha$ & $\sigma$ & 
$\log_{10}\left(\frac{M_{\rm sub}}{M_{\rm host}}\right)$\\
& [$\Mass$]                &               &          &          &   \\
\hline
\hline
   &$  9.2\times10^{12}$ & $-2.05$ & $-0.87$ & $0.17$ & $-1.8$ \\
   &$  2.7\times10^{13}$ & $-2.06$ & $-0.89$ & $0.16$  & $-2.5$\\
MS &$  7.9\times10^{13}$ & $-1.98$ & $-0.88$ & $0.13$ & $-2.8$\\
   &$  2.3\times10^{14}$ & $-2.00$ & $-0.90$ & $0.10$  & $-3.5$\\
   &$  6.8\times10^{14}$ & $-1.86$ & $-0.87$ & $0.06$ & $-3.8$\\
\hline 
   &$  3.1\times10^{12}$ & $-2.00$ & $-0.83$ & $0.16$ & $-2.5$\\
HS &$  9.2\times10^{12}$ & $-2.05$ & $-0.88$ & $0.17$ & $-2.8$\\
   &$  2.7\times10^{13}$ & $-2.15$ & $-0.93$ & $0.14$ & $-3.5$\\
%
\hline
\end{tabular}
\caption[The best-fit parameters to the mass function of subhaloes residing
in haloes of different mass at $z=0$, using Eq.~\ref{sub:eq:submf}]{
The best-fit parameters to the mass function of subhaloes residing 
in haloes of different mass at $z=0$, using Eq.~\ref{sub:eq:submf}. The columns 
are as follows: (1) The N-body simulation from which the halo sample was
extracted. (2) The mean mass of the host haloes. (3) The logarithm of 
the amplitude. (4) The power-law index. (5) The damping strength. (6) The
minimum fractional subhalo mass included in the fitting. 
}
\label{sub:tab:params}
\end{center}
\end{table}

We have also fitted Eq.~\ref{sub:eq:submf} to the subhalo mass functions
in each halo mass bin, this time letting the slope $\alpha$ vary; we
list the best-fit parameters for $z=0$ in Table~\ref{sub:tab:params}. At
the low fractional mass end, where the subhalo mass function behaves
as a power-law, we generally find slopes that are lower than the
critical value, $\alpha=-1$ (which separates divergent from convergent
mass functions). The slopes we find are in broad agreement with
previous estimates of the power-law index of the subhalo mass
function, which range from $-0.8$ to $-1.0$ 
\citep{Moore1999, Ghigna2000, deLucia2004, Gao2004, Diemand2004,
Shaw2007, Diemand2007}. In particular, our results agree with those
from the much higher resolution simulations of individual galactic
halos of \cite{Springel08}, but are inconsistent with the steeper slope
advocated, also for galactic halos, by \cite{Diemand08}.

At the high mass end, the subhalo mass function departs from a
power-law and decreases exponentially. This behaviour was previously
detected in N-body simulations (at lower significance) by
\cite{Giocoli2008b} (and predicted analytically by 
\citealt{vandenBosch2005}).
However, this feature was not apparent in earlier
studies which used resimulations of individual haloes. Resimulations
of single objects have the advantage that computational effort can be
focused. A halo can be resolved with a vast number of particles and
its substructures identified over a large range of
masses. Unfortunately, this approach comes at the price of losing the
rich information contained in the variety of assembly histories,
relaxation states and, more importantly, the population of high mass
subhaloes. As can be seen from Fig.~\ref{sub:fig:submf}, the abundance of
these objects is much lower than that of smaller subhaloes -- usually
we would find just a few in each halo. Because these halos are so
rare, the damping of the power-law at high $\fmass$ is missed in
individual resimulations. By contrast, with the huge sample of haloes
and their massive subhaloes in our analysis, we can robustly probe
this subhalo mass range. 

Even though the subhalo mass function appears roughly universal (e.g.
\citealt{Moore1999}), we have detected at every redshift a small
dependence on the mass of the host system. Small substructures of the
same fractional mass are more abundant in high mass haloes than in low
mass haloes. This correlation has also been seen in a number of other
studies \citep[e.g.][]{Gao2004,Shaw2007,Diemand2007}. However, we also
find evidence that this trend holds only in the power-law region of
the subalo mass function and actually reverses at the high mass end -
low mass haloes seem to host relatively more massive subhaloes than do
high mass haloes.

Perhaps surprisingly, the variety of features present in the mass
function of subhaloes is consistent with a relatively simple picture.
There are two key ingredients that shape the subhalo mass function:
(i) the mass function of infalling objects and (ii) the dynamical
evolution of subhaloes orbiting within the host halo due to dynamical
friction and tidal stripping.  The first of these is responsible for
the universality described above and sets the subhalo mass function to
first order. As first found by \citet{LaceyCole1993} using the
extended Press Schechter formalism, and confirmed by
\cite{Giocoli2008b} using N-body simulations, the mass function of
subhaloes at infall is almost independent of host halo mass and
redshift when expressed as a function of $M_{\rm sub}/M_{\rm host}$,
and can be described as a power-law with a high mass cut-off.

After subhaloes fall into a host halo, their orbits sink due to
dynamical friction and, at the same time, the subhaloes lose mass due to tidal
stripping. These processes cause the subhalo mass function to evolve
away from its form at infall. The rates for these processes depend on
the fractional mass of the subhalo, $M_{\rm sub}/M_{\rm host}$, and on the dynamical
timescale of the host halo. Therefore, if all haloes had identical
structure and assembly histories, these processes would preserve a universal form
for the subhalo mass function, independently of $M_{\rm host}$. However,
haloes of different masses on average assemble at different redshifts 
in spite of the similar mass function of subhaloes at infall, and this
breaks the universal shape of the subhalo mass function, as discussed
by \cite{vandenBosch2005} and \cite{Giocoli2008b}.  On average,
massive haloes are younger than their less massive counterparts and
they are more likely to have experienced recent mergers
\citep{LaceyCole1993}. These provide a fresh source of substructures
which have had less time for orbit decay due to dynamical friction and
to be tidally stripped. High mass haloes are therefore expected to
have more substructures than low mass haloes. Another effect which
acts in the same direction is that small haloes tend to accrete their
subhaloes at higher redshifts when dynamical timescales are
shorter. As a result, they strip out mass from the substructures more
quickly than large haloes, where massive substructures can survive for
longer.  

\subsection{Most massive subhaloes}

\begin{figure} 
\begin{center}
\includegraphics[width=8.5cm]{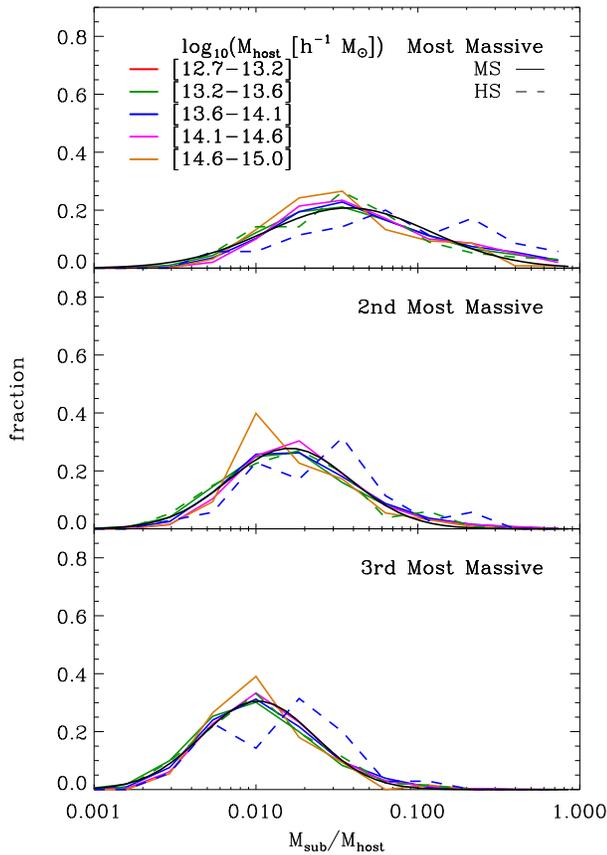}
\caption[The distribution of the fractional mass of the 1st, 2nd and 3rd
largest substructures in haloes of different mass at $z=0$.]{The distribution of
the fractional mass, $\fmass$, of the 1st, 2nd and 3rd largest substructures in
haloes of different mass at $z=0$. The solid lines show the results from the MS
while the dashed lines show the results from the HS.  In each panel, the black
solid lines indicate the log-normal function that best fits our results. Note
that only substructures resolved with $20$ particles or more are displayed.
\label{sub:fig:mm}}
\end{center}
\end{figure}

The high-mass tail of the distribution of substructure is examined in greater
detail in Fig.~\ref{sub:fig:mm}. The three panels in this plot display the
distribution of the fractional mass, $\fmass$, for the first, second and third
largest substructures within haloes of different mass at $z=0$. As before,
results from the MS and HS agree very well. 

In contrast to the results presented in the previous subsection, the
distributions of fractional masses seem to be independent of the host halo
mass. (We have also checked that they are independent of redshift.)  In
particular, in every halo, the fractional masses follow a log-normal
distribution with mean $\langle \log_{10} (\fmass) \rangle = -1.42$, $-1.79$
and $-1.99$, and standard deviation $\sigma_{\log_{10}(\fmass)} = 0.517$,
$0.382$ and $0.348$ for the each of the three largest subhaloes respectively.
Albeit with considerable scatter, these values imply that the most massive
substructure contains typically $3.7\%$ of the total mass of the halo while the
second and third most massive subhaloes contain $1.6\%$ and $1\%$ of the mass
respectively.

Due to the large dispersions, the distributions can only be measured reliably
in haloes resolved with a large number of particles. For instance, the mean
fractional mass of subhaloes is overestimated for haloes resolved with fewer
than $\sim 1000$ particles (the exact limit depends on the scatter and mean of
the true distribution), i.e. $\sim 1\times10^{12}\,\Mass$ in the MS and $\sim
1\times10^{11}\,\Mass$ in the HS. The upward bias is caused by the finite
resolution of the simulations (there is a limit on the smallest subhalo that we
can identify) which truncates the low mass tail of the distribution of
fractional masses. 

Hints of a universal behaviour of the fractional masses of the largest
subhaloes were already detected by \cite{deLucia2004} (although they claim a
weak dependence with host halo mass). Our results are broadly consistent with
theirs but, with the large halo catalogues from the MS and HS, we are able to
probe the full probability distribution function robustly.

The apparently universal shape of these distributions could, in principle, be
understood within the broad picture just discussed.  Presumably it reflects the
distribution of masses of the infalling haloes which, as we have seen, is
independent of the host halo mass \citep{LaceyCole1993,Giocoli2008b}. The large
scatter must then result from the large range of accretion histories at a given
host halo mass. We leave further investigation of these ideas for future work.


\begin{figure*} 
\begin{center}
\includegraphics[width=17cm]{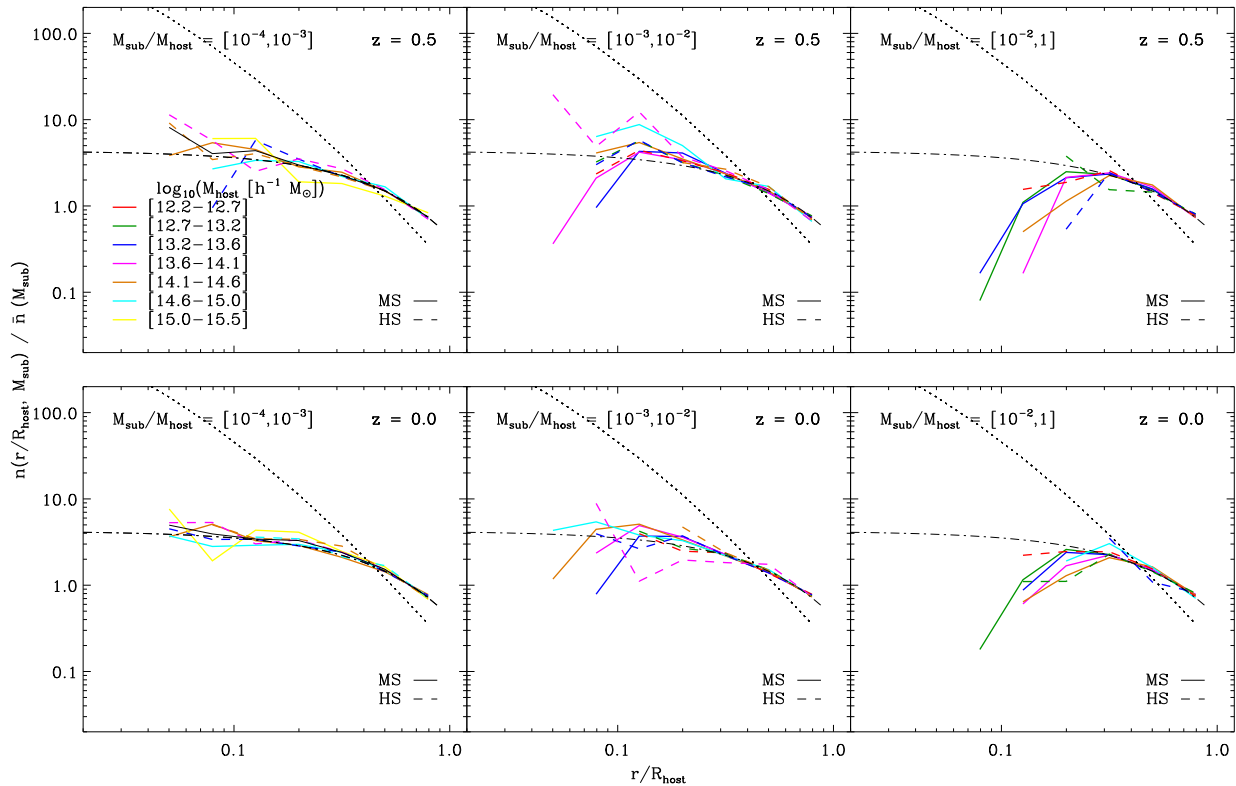}
\caption[The number density of subhaloes relative to the mean within
$\rhost$, as a function of the distance to the centre of their host halo, in
units of the radius of the host halo.]{The number density of subhaloes relative
to the mean within $\rhost$, as a function of the distance to the centre of
their host halo, in units of the radius of the host halo, $\rhost$. Each
column corresponds to a different fractional mass range for subhaloes, while
the rows display the results at two separate redshifts.  The solid and dashed
lines show the density profiles from subhaloes in the MS and HS respectively.
In both cases, the different colours correspond to subhaloes residing in haloes
of different mass as shown in the legend. The black dotted lines in each panel
indicate the mean dark matter density profile of haloes in our simulations.
Results are shown only for subhaloes resolved with at least 200 particles. The
top row shows results for $z=0.5$ and the bottom row for $z=0$.  
\label{sub:fig:rad_dist}}
\end{center}
\end{figure*}


\subsection{Radial distribution of subhaloes} \label{sub:sec:rad}

Fig.~\ref{sub:fig:rad_dist} shows the number density of subhaloes as a function
of radius, relative to the mean number density of substructures within
$r_{200}$ in the same fractional mass range. Each panel focuses on
substructures of different masses, from small subhaloes ($10^{-4} < \fmass <
10^{-3}$) in the leftmost panel to large ones ($10^{-2} < \fmass < 1$) in the
rightmost panel. As in previous plots, lines of different colours show results
for subhaloes that reside in haloes of different mass, and the different line
types (solid and dashed) indicate the results for the two simulations. We also
plot the radial profile of the dark matter as a black dotted line in each
panel.

Comparison of the MS and HS indicates that our results are insensitive to the
mass resolution (although the overlap between the two simulations is limited).
As in previous studies (e.g. \citealt{Gao2004}), we find that the radial
distribution has little dependence on the host halo mass at a given $\fmass$.
This is quite remarkable since each panel mixes subhaloes that: (i) are
resolved by numbers of particles that differ by orders of magnitude and (ii)
occupy haloes which are in a variety of dynamical states (age, relaxation,
etc). We also see that in all cases, the radial distribution of subhaloes is
less centrally concentrated than the dark matter, as was also found in previous
studies \citep[e.g.][]{Ghigna1998, Ghigna2000,Gao2004,Diemand2004,Nagai2005,Shaw2007,Springel08}. 

In addition, we see a significant difference between the distribution of
massive subhaloes ($M_{\rm sub} > 10^{-2}\,\Mhost$) and that of small ones
($M_{\rm sub} < 10^{-3}\,\Mhost$).  While the overall radial profiles seem to
be fairly independent of subhalo mass, the more massive subhalos tend to avoid
the central regions of the host halo, while the less massive ones have a more
centrally concentrated distribution \citep[see also][]{deLucia2004}. However,
the distributions agree in the outer parts of the halo. \cite{Springel08} 
found a similar effect to ours in the Aquarius set of simulations of galactic 
halos which, although limited in number, span a huge dynamic range in subhalo mass.

These dissimilar density profiles for different subhalo masses have a
simple dynamical explanation \citep[e.g.][]{Tormen1998,Nagai2005}.
Once a halo falls into a more massive system, dynamical friction and
tidal striping start to act. The accreted subhalo will rapidly be
stripped of its outer layers and will lose a significant fraction of
its mass during the first pericentric passage. This mechanism
naturally differentiates the radial distribution of substructures of
different masses: massive structures sink more rapidly due to
dynamical friction and, as a result, also lose mass more quickly by
tidal stripping.  Therefore they do not survive long in the central
regions, in contrast to small subhaloes.  The massive subhaloes which
are present in the halo must have been accreted more recently than the
average low mass subhalo. The timescale for dynamical friction depends
on the relative mass of the subhalo and its host halo, not on their
absolute values, which would explain the approximate independence of
the distribution on the host halo mass.

\subsection{Angular distribution of subhaloes}

\begin{figure} 
\begin{center}
\includegraphics[width=8.5cm]{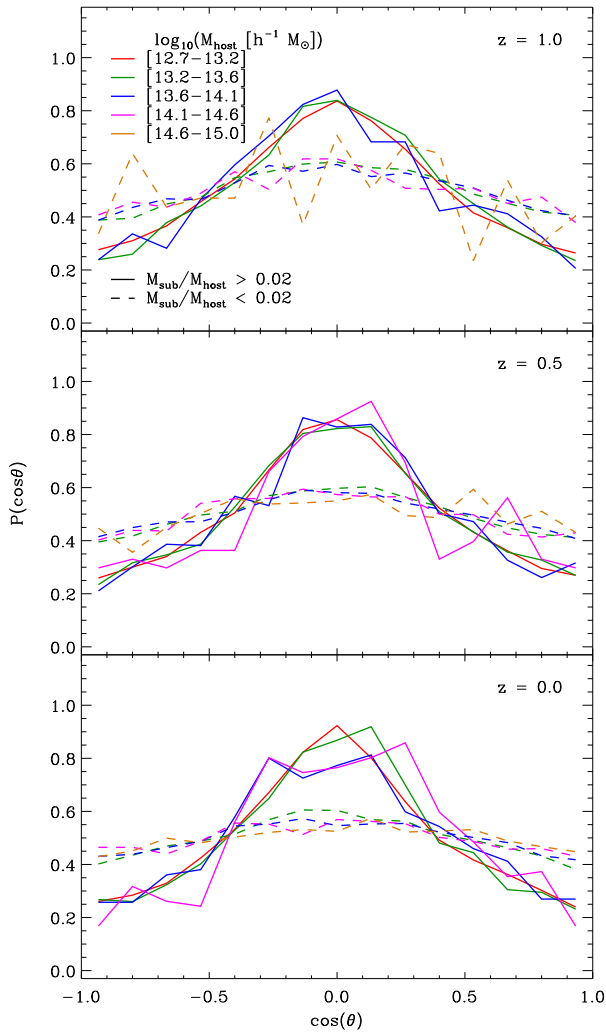}
\caption[The probability density distribution of the cosine of the angle
between the angular momentum vector of the host halo and the vector joining its
centre with that of the subhalo.]{ The probability density distribution of the
cosine of the angle $\theta$ between the angular momentum vector of the host
halo and the vector joining its centre with that of the subhalo. Each panel 
shows results for a different redshift, as indicated by the label. The lines in
each panel display the distribution for subhaloes in two different mass bins:
$0.02 < \fmass < 1$ (solid lines) and $0.004 < \fmass < 0.02$ (dashed lines).
Lines of different colour indicate subhaloes residing in host haloes of
different masses, as shown in the legend. An isotropic angular distribution
corresponds to a horizontal line.
\label{sub:fig:angle}}
\end{center}
\end{figure}

To end this section we investigate the angular distribution of subhaloes within
dark matter haloes. Previous work has examined the relationship between the
angular distribution of substructures and the {\em shape} of the host halo
\citep{Tormen1997,Libeskind2007,Knebe2008b,Knebe2008a}. Here, we examine
instead the orientation relative to the {\em spin axis} of the host halo.
Fig.~\ref{sub:fig:angle} shows the probability distribution function of the
cosine of the angle between the angular momentum vector of the host halo and
the vector joining its centre with that of the subhalo. 
We show results for two separate ranges of subhalo mass: subhaloes with mass smaller than $2\%$
of the host halo mass (dashed lines) and those with masses greater than $2\%$
(solid lines). We distinguish different host halo masses by different colours,
and show different redshifts in different panels.
Note that we only display results for the MS simulation for clarity.

As shown by Bett et~al. (2007), the accuracy of the measurement of spin
direction in the MS degrades significantly (uncertainty $> 15\deg$) for haloes
resolved with fewer than $1000$ particles or for those where the spin
magnitude, $|j|$, is such that:

\begin{equation}
\frac{|j|}{ \sqrt{G\,\Mhost\,\rhost }} < 10^{-1.4},
\end{equation}

\noindent where $G$ is Newton's gravitational constant. Although the inclusion
of haloes that do not satisfy these criteria does not seem to affect our
results quantitatively, we have chosen to show only those haloes that met these
requirements, so that the angle relative to the spin axis can be reliably
determined.

We see from Fig.~\ref{sub:fig:angle} that the angular distribution of subhaloes
tends to be aligned perpendicular to the spin axis of the host halo. (We remind
the reader that in this plot, an isotropic angular distribution would
correspond to a horizontal line, while a distribution aligned at $90\deg$ to
the spin axis will peak around $\cos \theta \sim 0$.) The strength of this
alignment effect depends on the fractional subhalo mass, $\fmass$, being much
stronger for higher mass subhaloes. We also see that the angular distribution
for a given $\fmass$ is almost independent of the host halo mass and the
redshift (see also \citealt{Kang2007}).

We can understand this behaviour qualitatively as reflecting the growth of
haloes by the accretion of dark matter (in halos or more diffuse form) along
filaments. The central regions of haloes acquire most of their angular momentum
at a relatively late stage from the orbital angular momentum of this infalling
material, and so will tend to have spin axes perpendicular to the current
filament \citep[e.g][]{Shaw2006,Aragon2007}. On the other hand, insofar as the subhaloes
``remember'' the direction from which they fell in once they are orbiting
inside the host halo, then their spatial distribution will tend to be aligned
with the filament from which they were accreted, and so will be perpendicular
to the spin axis. We can also understand the dependence of the strength of this
alignment on subhalo mass in this picture. Subhaloes with large $\fmass$ on
average have been orbiting in the host halo for less time than haloes of lower
$\fmass$, due to the combined effects of dynamical friction (which causes
higher mass subhaloes to sink faster) and tidal stripping (which converts
high-mass subhaloes to low mass). We expect subhaloes increasingly to lose
memory of their initial infall direction the longer they have orbited in the
host halo (which in general is lumpy and triaxial). Since high $\fmass$
subhaloes have undergone fewer orbits, their current angular distribution
should be more closely aligned with their infall direction, and therefore with
the current filament, compared to subhaloes of lower mass. 

Our results seem generally consistent with previous simulation results on the
alignment of the subhalo distribution with the shape of the host halo, and the
relationship between the shapes and the spin axes of halos. \citet{Tormen1997}
found that the angular distribution of subhaloes as they fall into a host halo
(crossing through $r_{200}$) is anisotropic, and tends to be aligned along the
major axis of the host halo. Previous studies \citep[e.g.][]{Knebe2004,Zentner2005,Libeskind2007} found
that the angular distribution of subhaloes within a host halo is aligned along
the major axis of the host halo. On the other hand, \citet{Bett2007} showed
that the angular momentum of a halo is generally aligned with its minor axis
and perpendicular to its major axis. Putting these results together, we would
expect the subhalo distribution to be aligned perpendicular to the spin axis of
the host halo, but ours is the first study to demonstrate this directly, and
also to demonstrate that the strength of the alignment depends on subhalo mass.
\section{Mergers between subhaloes} \label{sub:sec:merger}

\begin{figure*} 
\begin{center}
\includegraphics[width=17cm]{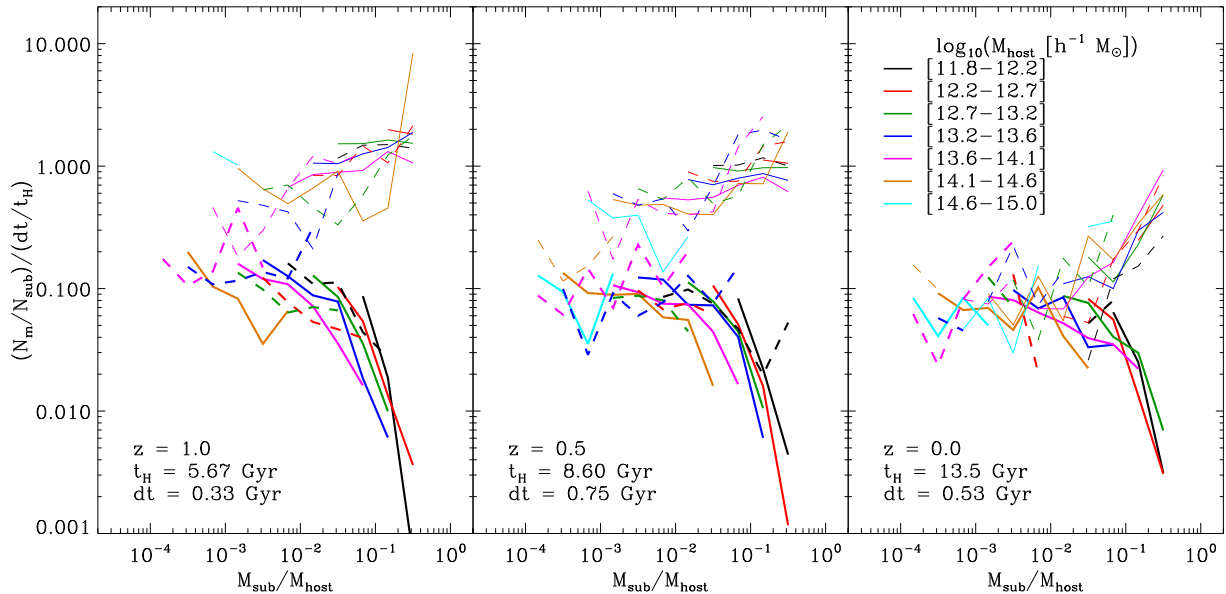}
\caption[The mean number of satellite mergers per subhalo and per unit of time
relative to the age of the universe, as a function of the mass of the
progenitor of the less massive object involved in the merger.]{ The mean number
of satellite mergers per subhalo and per unit of time relative to the age of
the universe, as a function of the mass of the progenitor of the less massive
object involved in the merger. Two cases are displayed: the number of
satellites destroyed or merging with the main substructure (top thin lines) and
the number of mergers between two satellites (thick bottom lines). The solid
lines show results from the MS while the dashed lines show results from the HS.
As indicated by the legend, in both cases, coloured lines represent results for
haloes of different mass. The three panels are for three different redshifts:
$z=1.0$, $z=0.5$ and $z=0$. Note that in the case of the merger between a
satellite and a central structure, we show examples involving subhaloes of at
least 200 particles, but we reduce the limit to 50 particles in the case of
mergers between two satellites. In each panel, the legend states the redshift,
the age of the universe, $t_{\rm H}$, and the time interval, ${\rm d} t$, over
which we measure the rates.
\label{sub:fig:subsub}}
\end{center}
\end{figure*}
 
\begin{figure} 
\begin{center}
\includegraphics[width=8.5cm]{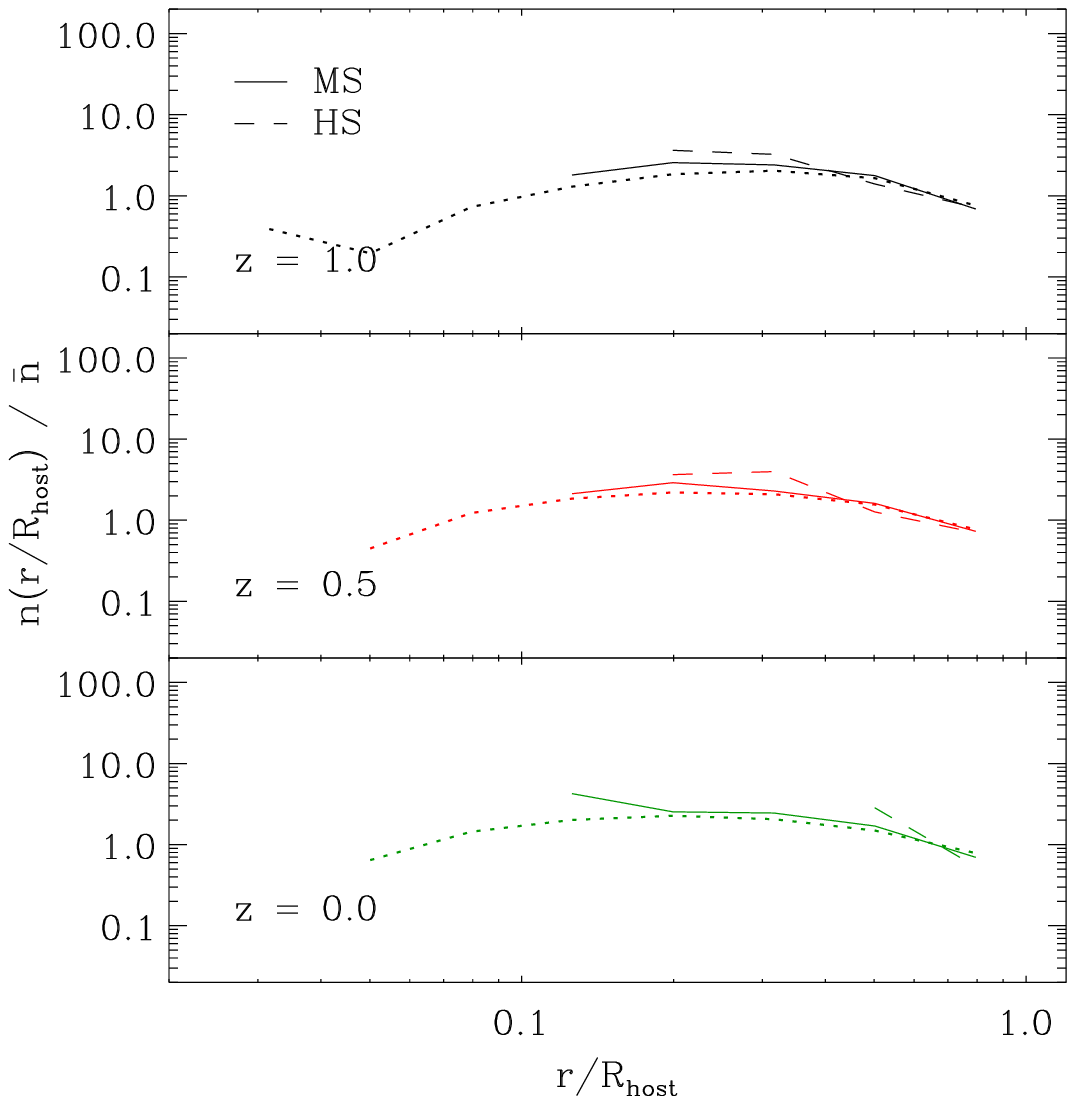}
\caption[The number density of subhalo-subhalo mergers relative to the mean
density of subhaloes within $r_{200}$ as a function of the distance to the
centre of the host halo.]{ The number density of subhalo-subhalo mergers
relative to the mean density of subhaloes within $r_{200}$ as a function of the
distance to the centre of the host halo. The results from the MS are shown by
solid lines while the results from the HS are shown by dashed lines. In each
subpanel the dotted lines show the radial distribution of all subhaloes
(regardless of whether they are merging or not) in the MS. Mergers involving
subhaloes resolved with at least 50 particles are included in the plot.
\label{sub:fig:r_dist}}
\end{center}
\end{figure}

As we have seen, once a halo is accreted by a larger one, its outer layers are
rapidly stripped by tidal forces. However, the core generally survives the
accretion event and can still be recognized as a substructure or satellite
subhalo within the host halo for some time afterwards. Furthermore, not only
may the main infalling halo survive, but also substructures within it. In this
case, there are substructures inside substructures.

While orbiting inside the halo, dynamical friction causes the orbit of a
subhalo to lose energy and to sink towards the centre of the host halo. As the
subhalo sinks, it suffers further tidal stripping. Eventually, the subhalo may
be totally disrupted: there is a merger between the satellite subhalo and the
central subhalo.  Nevertheless, on its way to destruction, a subhalo can
survive for several orbits during which it may experience a merger with another
satellite subhalo.  In the following subsections we will investigate the
merging of these substructures.

The interaction between subhaloes was previously investigated in cosmological
simulations by \citet{Tormen1998}, who studied the rate of penetrating
encounters between satellite subhaloes, but not the merger rate.
\citet{Makino1997} derived an expression for the merger rate between subhaloes
in galaxy clusters based on an entirely different approach, motivated by the
kinetic theory of gases. In this case, the merger rate per unit volume between
halos of mass $M_1$ and $M_2$ is $R_{\rm merge} = n_1 n_2
\sigma(v_{12})v_{12}$, where $n_1$ and $n_2$ are the respective number
densities, $v_{12}$ is the relative velocity, and $\sigma(v_{12})$ is the
merger cross-section. They used N-body simulations of isolated spherical halos
to derive merger cross-sections for equal-mass halos as a function of their
relative velocity, and then assumed that mergers in clusters occurred between
pairs of subhaloes drawn from random uncorrelated orbits, with a Maxwellian
distribution of relative velocities. The \citeauthor{Makino1997} expression was
then extrapolated to the case of unequal subhalo masses and incorporated into a
semi-analytical model of galaxy formation by \citet{Somerville1999} and
\citet{Hatton2003}. We will investigate below whether the \citet{Makino1997}
kinetic theory approach has any applicability to subhalo mergers in a realistic
cosmological context.

\subsection{Subhalo merger rate}

Fig.~\ref{sub:fig:subsub} shows the mean merger rate of satellite subhaloes,
plotted against the fractional mass of its progenitor. This is the mass of the
satellite before accretion divided by the mass of the host halo at the time of
the merger. The rate is normalized per subhalo, with time in units of the age
of the universe at that redshift. This normalized rate is thus roughly equal to
the probability that a satellite subhalo will merge over one Hubble expansion
time. A rate higher than unity indicates that the process happens on a
timescale shorter than a Hubble time. There are two sets of curves in this
figure: (i) the thinner, higher amplitude lines which show mergers between a
satellite and a central subhalo, as a function of the subhalo mass, and (ii)
the thick lines which correspond to satellite-satellite mergers, plotted as a
function of the mass of the smaller subhalo. As in previous plots, different
line colours show different host halo masses, and different line styles (solid
and dashed) show the two simulations used.

We see from Fig.~\ref{sub:fig:subsub} that over most of the subhalo mass range
resolved by our simulations (for $\fmass \gsim 10^{-3}$), it is more likely for
a satellite subhalo to merge with the central subhalo than with another {\it
more massive} satellite subhalo.  For instance, at $z=1$, taking into account
all host haloes, there are $17155$ satellites which merge with a central
subhalo over one timestep, while the number of satellites involved in a merger
with another satellite over the same period is $509$, a ratio of ~$40:1$. The
situation is similar at $z=0$ even though the ratio decreases to $6:1$ ($1645$
{\em vs} $290$). In general, the likelihood of both merger rates slightly
decreases at lower redshifts. This may reflect the slower build-up of structure
(relative to the Hubble time) as the universe becomes dominated by vacuum
energy.

As we consider smaller subhalo masses, we see a decrease in the destruction
rate (see the appendix for a discussion of overmerging effects due to
insufficient mass resolution). This may be due to the inefficiency of dynamical
friction for low mass structures. On the other hand, there is an increase in
the satellite-satellite merger rate as the subhalo mass decreases. Presumably
this is due to the increasing number of potential merger partners, reflecting
the form of the subhalo mass function. Additionally, the abundance of both
types of mergers is similar in the range $10^{-3} < \fmass < 10^{-2}$.
Unfortunately, at this point our results from low mass haloes become limited by
resolution (i.e. we cannot identify smaller substructures) and the results from
high mass haloes become dominated by Poisson noise (i.e. less than one merger
event in the whole simulation). Over the range that is reliably covered, we can
see no strong systematic differences in Fig.~\ref{sub:fig:subsub} between the
results derived from host haloes of different masses. This agreement is quite
remarkable given the relatively large dynamical range resolved in the
simulations.

A merger between two objects is not always a straightforward quantity to define
in numerical simulations. The problem originates from the fact that any
definition is intrinsically linked to the mass and time resolution of the
simulation. For instance, if in a higher resolution simulation we identify the
remnant of a subhalo down to a smaller mass threshold, then the mass ratio of
the merger, as well as the time at which it happens, could, in principle,
disagree with the values measured in a lower resolution simulation. Similarly,
with better time resolution, one could follow the mass loss of a subhalo more
accurately which, in principle, could also change the measured mass ratio of
the merger. To avoid these problems, we have chosen to use in
Fig.~\ref{sub:fig:subsub} the mass of the satellite before accretion, rather
than the mass at the moment of the merger. 

For all these reasons it is very important to note the agreement in
Fig.~\ref{sub:fig:subsub} between the results from the MS (solid lines) and
those from the HS (dashed lines). This agreement gives us confidence that our
results are not sensitive to mass resolution. (Note that this is not true for
subhaloes resolved with fewer particles as shown in the appendix.) Furthermore,
the weak dependence of the quantities plotted in Fig.~\ref{sub:fig:subsub} on
host halo mass confirms this conclusion. In practice, a subhalo of $\fmass =
0.1$ in a host of $10^{12}\,\Mass$ exhibits the same behaviour as a subhalo of
the same fractional mass but in a halo of $10^{14}\,\Mass$ even though the
latter is resolved with $100$ times more particles. This is quite remarkable.

One of the reasons for the insensitivity to mass resolution comes from our
definition of a merger (see \S \ref{sub:sec:method:haloes}). We do not tag an
event as a merger when we cannot identify the subhalo anymore, but rather when
it has lost a fixed fraction of its most bound mass. This definition responds
more to dynamical processes than to numerical ones.

The implications of discrete time measurements are less clear for our
definition of a merger. As an example, consider the case of very poor time
resolution, and a halo that is just about to fall into a larger one. If tidal
forces stripped off more than $95$\% of its mass before the next snapshot, then
we would have identified this event as a merger. On the other hand, if the time
resolution were good enough, we could have identified the subhalo at
intermediate stages, updating its mass and the corresponding most bound 10
percent. As long as stripping does not occur on a timescale much shorter than
the time resolution, it is even possible to imagine that the line of
descendants continues indefinitely. However, since a merger is not a discrete
event, better time resolution does not necessarily imply a more accurate
determination of a merger. With infinite time resolution, we would follow most
of the merging process down to the point when mass resolution becomes
important, i.e. every subhalo disruption would be caused by lack of mass
resolution.

However, the typical timescale for dynamical friction and tidal disruption is
$T_{\rm fric} \sim t_H$ for $\fmass \sim 0.1-0.2$ \citep{Jiang2008}, i.e. much
longer than the time spacing of our simulation outputs ($\sim 300$Myr).
Furthermore, subhalo mergers seem to take place very fast. Both these factors
suggest that time resolution is not an issue for this study. In fact, we have
checked that our results do not change if we choose snapshots that are twice as
widely spaced as those used to build the merger trees. Nevertheless, we advise
the reader to keep these limitations in mind.

\subsection{Characterization of subhalo-subhalo mergers}

In most cases the subhalo-central merger occurs very close to the potential
minimum of the host halo. The spatial location of satellite-satellite mergers,
on the other hand, has a very distinctive distribution. In the following
subsection we investigate this further.

\subsubsection{Radial distribution of satellite-satellite mergers}

First, in Fig.~\ref{sub:fig:r_dist} we look at the spherically averaged radial
distribution of satellite-satellite mergers. The figure shows the number
density of mergers, relative to the mean density of subhaloes within $r_{200}$,
as a function of the distance to the centre of the halo. We also display, as
dotted lines, the distribution of all the substructures from the MS\footnote{At
first sight, the distribution of all substructures seems to disagree with the
results of Section~\ref{sub:sec:rad}. Since the subhalo population is dominated
by small mass objects, one would naively expect the distribution of all
substructures to follow that of the smallest subhaloes; as seen in
Fig.~\ref{sub:fig:rad_dist}, this has a slope which is always negative.
However, in practice, the dominant effect is the high abundance of low mass
host haloes in which only massive substructures can be resolved. As a result,
the distribution of all subhaloes in the MS resembles the distribution of the
most massive substructures.}.

At every redshift plotted, the radial distribution of mergers is proportional
to the radial distribution of subhaloes. This implies that most of the mergers
between subhaloes do indeed occur in the outer regions of the host halo. Note
that in these regions the background density is lower than in the inner
regions, making it easier to identify subhaloes. For this reason, we can follow
satellite-satellite mergers down to structures resolved with $50$ particles, as
opposed to the minimum of $200$ particles we require for central-satellite
mergers.

Our results do not appear consistent with the naive expectation from a gas
kinetic theory approach that the number density of mergers should be
proportional to the number density of subhalo pairs, i.e $R_{merge} \propto
n_{sub}^2$.  This discrepancy indicates that most of the satellite-satellite
mergers do not occur because of random encounters between two unrelated
substructures. We investigate this idea further in the following subsection,
where we look back at the orbits of the subhaloes that merge.

\subsubsection{Orbits of merging satellites}

\begin{figure} 
\begin{center}
\includegraphics[width=8.5cm]{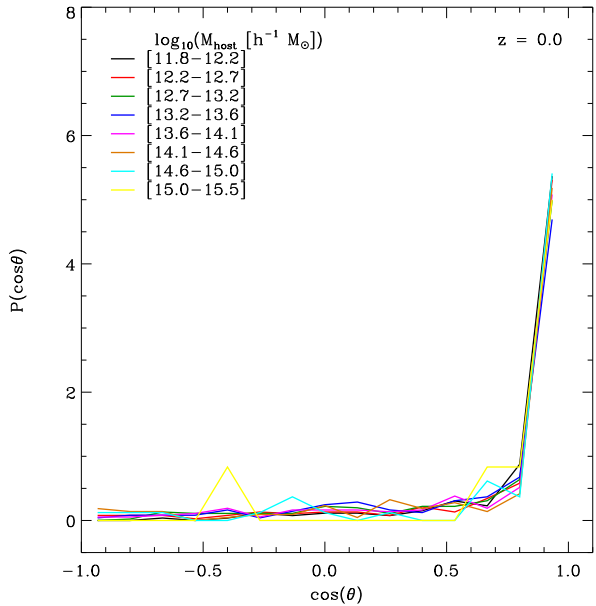} 
\caption[Probability distribution of the cosine of the separation angle
$\theta$ between the progenitors of two substructures that are going to
merge.]{ Probability distribution of the cosine of the separation angle
$\theta$ between the progenitors of two substructures that are going to merge.
The separation angle is measured at the last snapshot in which the subhaloes
were identified outside the halo that hosts the satellite-satellite merger.
Lines of different colours indicate mergers happening in haloes of different
masses as indicated in the legend. 
\label{sub:fig:mangle}}
\end{center}
\end{figure}

\begin{figure} 
\begin{center}
\includegraphics[width=8.5cm]{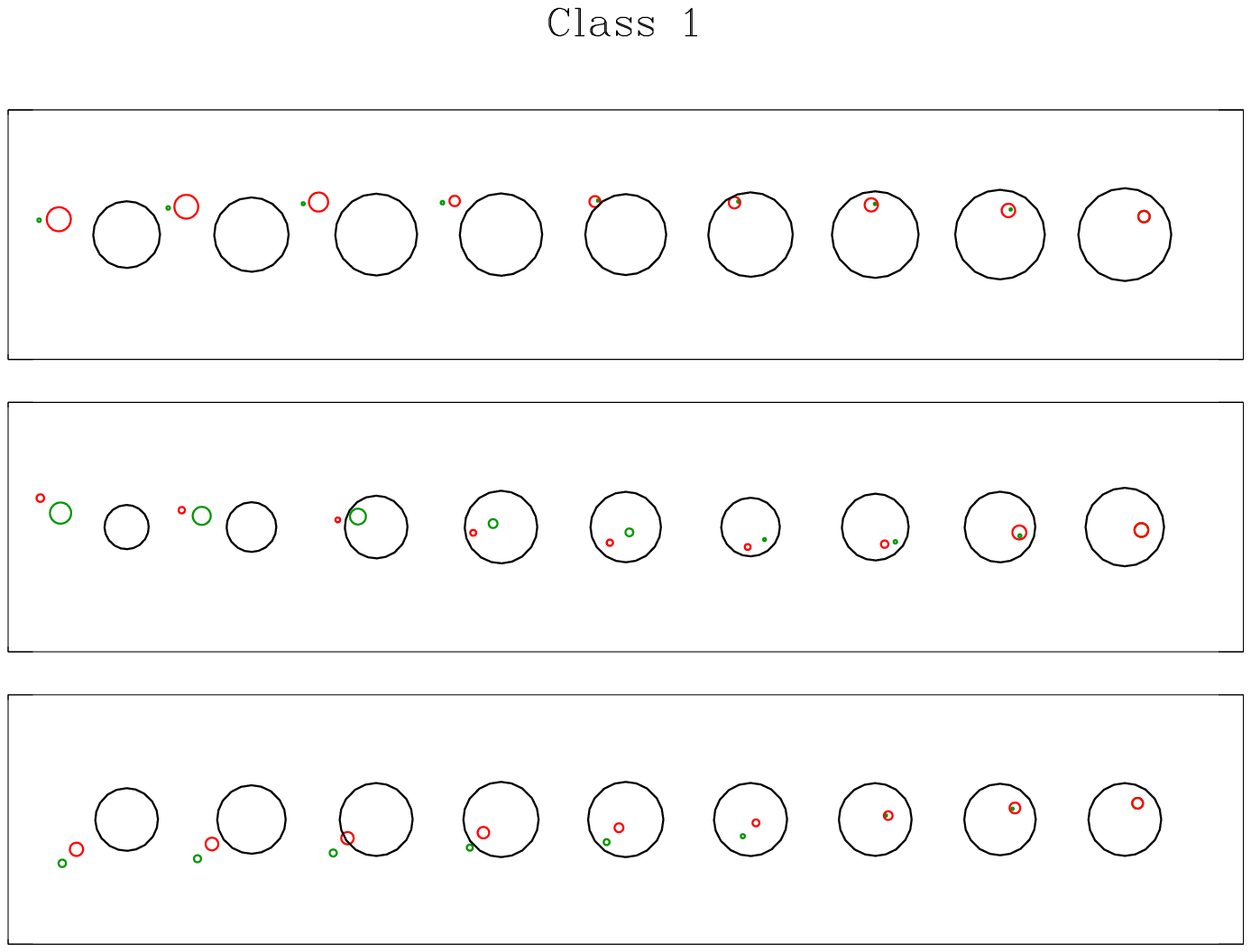}\\
\includegraphics[width=8.5cm]{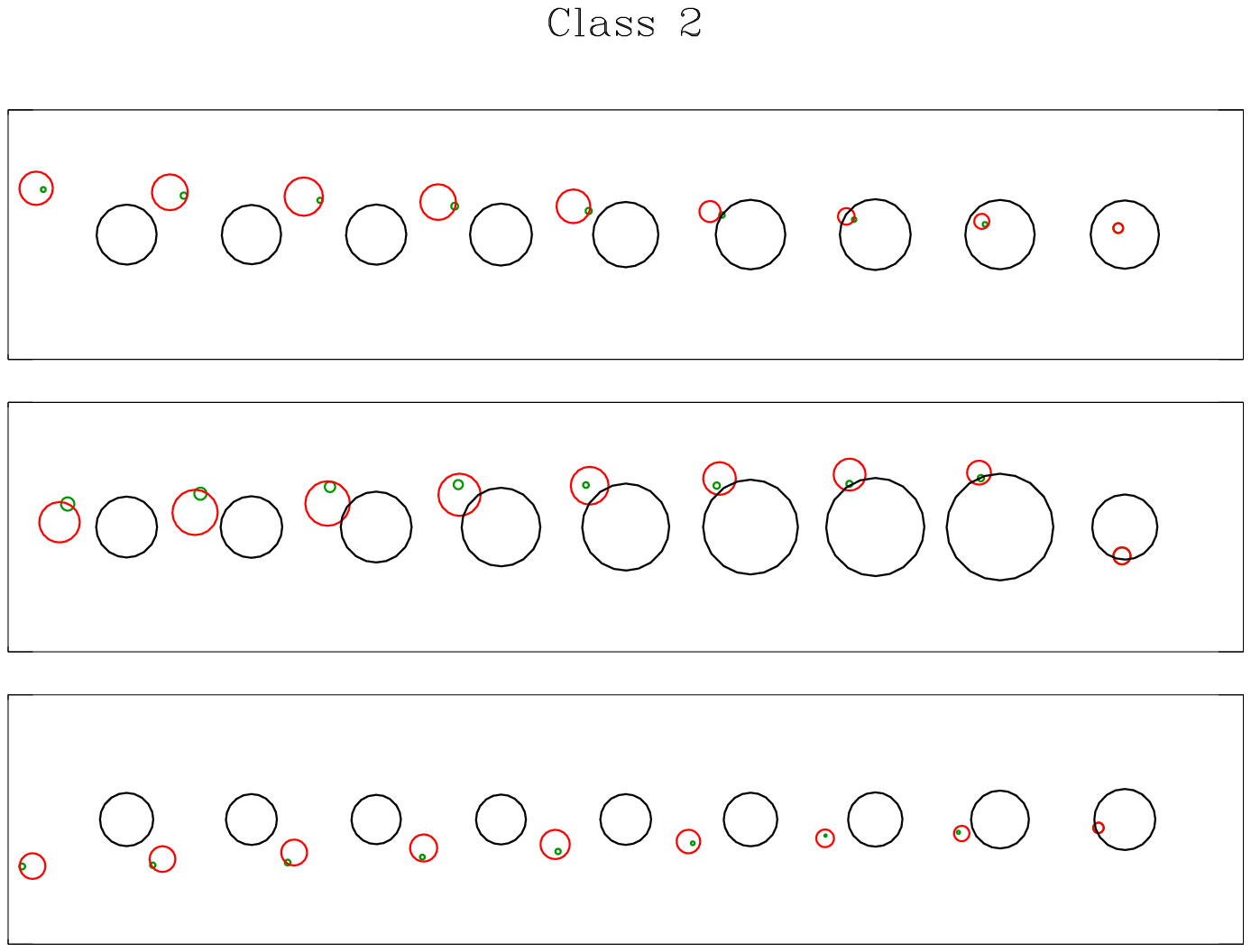}
\caption[Three representative examples extracted from the MS for each of the
two most common configurations between two satellite subhaloes that merge.]{
Three representative examples extracted from the MS for each of the two most
common configurations between two satellite subhaloes that merge. The plots
show the relative positions of the host and satellite halos in a time sequence,
with time increasing from left to right. The black circles correspond to the
halo that hosts the merger, while red and green circles show the positions of
the satellites involved in the merger.  The circles' radii are proportional to
the half mass radius of each substructure.  Class 1: in this case the
satellites were part of two separate haloes (red and green circles) before they
were accreted into a larger halo (black circles).  Class 2: both substructures
belonged to the same halo before it was accreted into the larger structure
which hosts the merger.  \label{sub:fig:example}}
\end{center}
\end{figure}

Fig.~\ref{sub:fig:mangle} shows the distribution function of the separation
angle $\theta$ between the progenitors of subhaloes involved in a merger. The
angle is measured from the centre of the host halo in which the merger is going
to take place, at the last snapshot in which both subhaloes were identified
outside the halo that later hosts the satellite-satellite merger. It thus
represents the angle between the subhaloes at the time they fall into the host
halo.  The first point to note is that the distribution seems to be universal
in the sense that it is roughly independent of the mass of the host halo. (We
have also checked that it is roughly independent of redshift.) However, the
most important feature is that the distribution is clearly dominated by small
separation angles. About $65$\% of the mergers occur between subhaloes that were
separated by less than $30\deg$ at the moment of accretion. (This percentage
increases to $73$\% for an angle of $43\deg$.) This demonstrates that the mergers
are mostly between two or more systems that were already dynamically associated
before they fell into the larger system. If the gas kinetic theory approach of
\citet{Makino1997} applied to this case, then the mergers would be between
subhaloes on random orbits, and we would expect a more uniform distribution in
$\cos\theta$. (It would not be completely uniform since the subhalo population
is not isotropic, as shown in Fig.~\ref{sub:fig:angle}.) 

More information about the orbits of merging subhaloes is given in
Fig.~\ref{sub:fig:example}, where we display three representative examples of
the two most common configurations of a satellite-satellite merger. These
examples correspond to real sequences found in the MS.  The plot tracks the
position of substructures up to the snapshot of the merger (which happens at
the rightmost position), starting on the left, $9$ snapshots earlier. We show as
a black circle the halo that hosts the satellite-satellite merger and, as green
and red circles, the progenitors of the subhaloes involved in the merger. The
red circle at the end of the sequence indicates the subhalo resulting from the
merger.  The radii of the circles are proportional to the half-mass radius of
the subhalo.

The two most common configurations are as follows: Class 1: the progenitors of
the subhaloes correspond to two separate haloes which were accreted at
approximately the same time. Note that, as shown by Fig.~\ref{sub:fig:mangle},
these haloes were spatially close at the time of accretion. Class 2: the merger
occurs between two substructures that were part of the same halo before it fell
into the host halo. In other words, there is a halo that contains two
substructures which survived the accretion and subsequently merged. The merger
event which started outside the main halo is completed inside it, as a
subhalo-subhalo merger.

Most subhalo mergers occur between substructures that are accreted close
together both in time and location.  Generally, they were already part of the
same system before it was accreted into a larger one, or were part of two
separate haloes that were about to merge.  This is probably a requisite for a
subhalo merger to occur. The potential generated by the other satellite has to
be at least comparable to that of the main halo. Hence, satellites accreted at
different angles will follow relatively independent dynamical histories and are
much less likely to merge.

\subsubsection{The mass ratio of subhalo mergers}

\begin{figure} 
\begin{center}
\includegraphics[width=8.5cm]{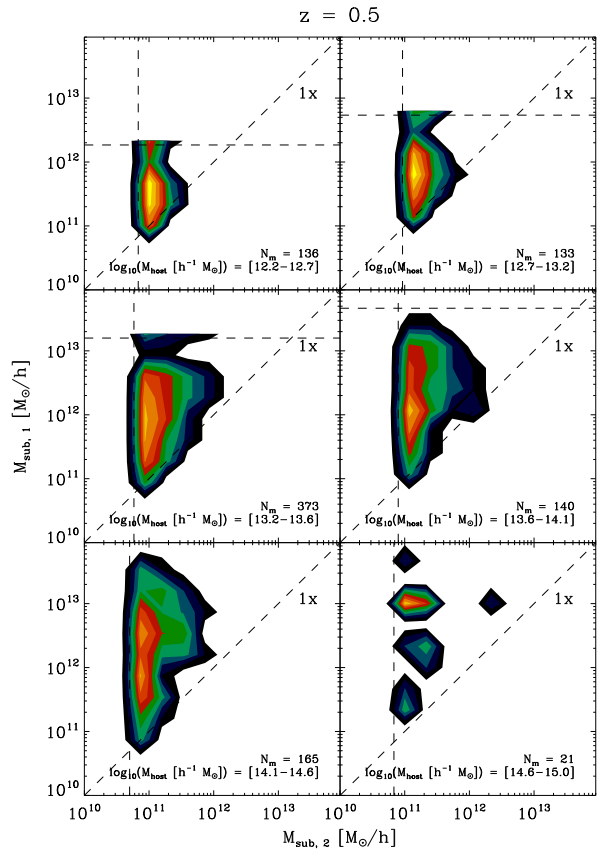}
\caption[Contour plot showing the logarithm of the number of
satellite-satellite mergers as a function of the masses of the merging
subhaloes at $z=0.5$.]{ Contour plot showing the logarithm of the number of
satellite-satellite mergers as a function of the masses of the merging
subhaloes at $z=0.5$. The $x$-axis indicates the mass of the smaller subhalo
while the $y$-axis indicates the mass of the larger subhalo.  The different
panels show the results for host haloes of different masses as indicated on
each panel. The numbers in the bottom right show the number of mergers
displayed in each panel.  The vertical dashed lines indicate the 200 particle
limit and the diagonal lines correspond to a 1:1 ratios between the masses of
the two subhaloes. The horizontal lines show the mass limit on the more massive
participant imposed by the choice of mass bin.\label{sub:fig:masses}}
\end{center}
\end{figure}

\begin{figure*} 
\begin{center}
\includegraphics[width=17cm]{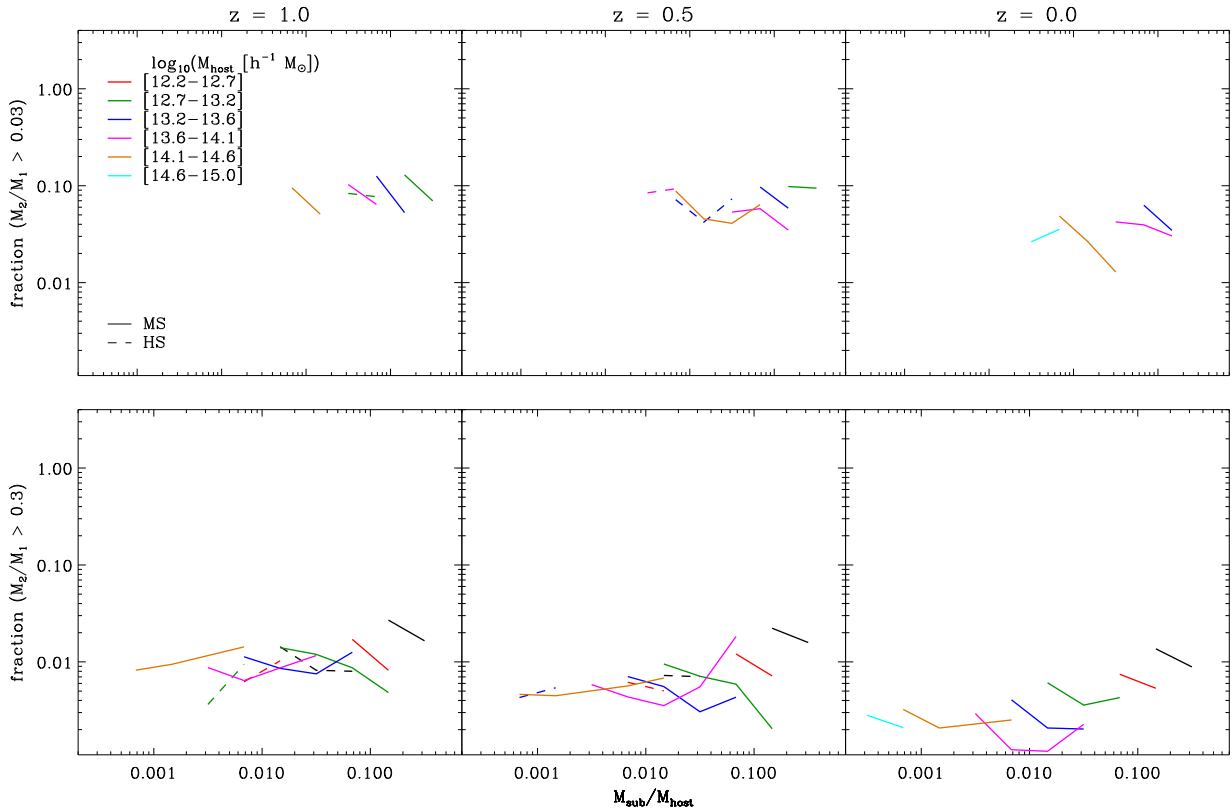} 
\caption[The fraction of substructures that have experienced a merger with
another substructure since the time of accretion into the current host halo.]{
The fraction of substructures that have experienced a merger with another
substructure since the time of accretion into the current host halo. The
$x$-axis gives the subhalo mass at the redshift shown, while the ratio
$M_2/M_1$ on the $y$-axis is for the two progenitors of the subhalo at the time
they merged. The results from the MS are shown by solid lines while the results
from the HS are shown by dashed lines. The coloured lines represent the data
from haloes of different masses, as indicated by the key. The two rows
correspond to different mass ratios between the subhalo progenitors involved in
the merger: $M_{\rm sub,2} > 0.03\,M_{\rm sub,1}$ (top row) and $M_{\rm sub,2}
> 0.3\,M_{\rm sub,1}$ (bottom row) where $M_{\rm sub,2}$ refers to the larger
satellite involved in the merger. The three panels display the results for
substructures identified at redshifts $z=1$, $0.5$ and $0$ respectively.
\label{sub:fig:fraction}}
\end{center}
\end{figure*}

In Fig.~\ref{sub:fig:masses}, we inspect the relative masses of the satellite
subhaloes which merge.  The $x$-axis indicates the mass of the smaller subhalo
and the $y$-axis shows the mass of the larger one. Interestingly, we find that,
for the range of host halo masses plotted, the most common merger is that
between two substructures of dissimilar masses, $M_{\rm sub,1} \sim 10\times
M_{\rm sub,2}$. Note that this trend is contrary to the naive expectation
whereby the mergers are simply proportional to the abundance of substructures,
in which case the maxima would be located around the line $M_{\rm sub, 1} =
M_{\rm sub,2 }$.  However, it is roughly consistent with the idea that
substructure mergers happen between two structures that were part of the same
halo before accretion. For instance, if the most common merger happens between
the main subhalo and its most massive substructure, then, as we have seen, we
would expect to find a mass ratio of 1:25 (see Fig.~\ref{sub:fig:mm}) and the
maxima of Fig.~\ref{sub:fig:masses} along $M_{\rm sub,1}\sim 10\times M_{\rm
sub,2}$

\subsection{Merger probability since accretion}

Finally, in Fig.~\ref{sub:fig:fraction} we plot the fraction of subhaloes at a
given redshift that have had a merger with another satellite subhalo since
accretion into the current host halo. The top panels show mergers between
satellites with a mass ratio greater than $0.03$, i.e. in which the less
massive subhalo has, at least, $3$\% of the mass of the larger one. In the bottom
panels we consider mergers between subhaloes with more similar masses: the
minimum mass ratio is $0.3$. 

The fraction of current subhaloes which have experienced a merger in the past
is a quantity strongly affected by resolution. For instance, in the history of
a subhalo resolved with $1000$ particles, because of our $200$ particle mass cut on
subhaloes, we can only record mergers with other subhaloes which account for at
least one fifth of the final subhalo mass. On the other hand, if our current
subhalo is resolved with $10000$ particles, then a much wider range of merger
mass ratios can be tracked. These considerations are further complicated by the
fact that we expect the measured mass of a subhalo to be less than the mass of
its progenitors at infall, due to tidal disruption and stripping; hence an
object that is below our $200$ particle limit at a particular redshift could have
been above this mass cut when it experienced the subhalo-subhalo merger.

To improve statistics, whilst at the same time attempting to avoid building a
resolution dependence into our results, we relax the particle number constraint
on subhaloes for this exercise. At the redshift a subhalo is identified (i.e.
the redshift plotted in Fig.~\ref{sub:fig:fraction}), we consider subhaloes of
$30$ particles or more. At the redshift of the subhalo merger, the progenitors
must both have $50$ particles or more to be counted. 

Fig.~\ref{sub:fig:fraction} shows that the probability of a subhalo merger is
constant for subhaloes of different mass. About $1\%$ percent of subhaloes have
had a merger with another subhalo with a mass ratio $>0.3$.  For a mass ratio
$>0.03$, this fraction increases to $\sim 10\%$. We also note that these
fractions show a weak decrease with redshift. 

\section{Summary and conclusions} \label{sub:sec:conc}

We have used the Millennium simulation, together with a simulation which has
$10$ times better resolution but about $100$ times smaller volume, to
investigate the general properties of the substructures within dark matter
haloes, including their merger rates.  Our main findings can be summarized as
follows:

In agreement with previous studies, we find that the mass function of low and
intermediate mass subhaloes follows roughly a power-law. However, we also find
an exponential cut-off in the mass function at high subhalo masses. We have
provided an expression, Eq.~\ref{sub:eq:submf}, that describes this behaviour
accurately.  We also detect a small but systematic dependence of the number of
subhaloes on the mass of the host halo. On average, at a given fractional mass,
$\fmass$, high mass haloes contain more low and intermediate mass substructures
than their less massive counterparts. In contrast, we find evidence that high
mass haloes contain fewer high mass subhaloes than do low mass haloes.  In
spite of this, the fractional mass of the first, second and third most massive
substructures is insensitive to the mass of the host halo and of the redshift.

We confirm that the radial and angular distributions of subhaloes are roughly
independent of the host halo mass and redshift. However, we find that the
radial distribution does depend on the subhalo mass relative to that of the
host halo. The subhalo distribution is less concentrated than the dark matter,
but the radial distribution of low mass subhaloes tends to be more concentrated
than that of high mass subhaloes. This difference can be understood as
resulting from the different efficiency of dynamical friction in subhaloes of
different mass. On the other hand, these discrepancies between the radial 
distributions of low and high mass subhaloes disappear in the outer parts
of the halo, as seen in recent ultra-high resolution simulations
of galactic halos \citep{Springel08}.



The angular distribution of subhaloes tends to be aligned perpendicular to the
spin axis of the host halo. This is probably due to an anisotropic mass
accretion - mergers happen preferentially along filaments. The alignment is
strong for the most massive subhaloes, but is much weaker for low mass
substructures since, on average, they have spent a few orbital times inside the
halo which would randomize their orientation.

We have found that satellite-satellite mergers do occur. Over most of
the mass range resolved in our simulations, they are subdominant when
compared with mergers between satellites and the central
subhalo. However, we see some indication that satellite-satellite
mergers are equally likely to satellite-central mergers for the lowest
mass subhaloes ($\fmass < 10^{-3}$). As for many other subhalo properties,
the merger rates appear to be a function of the fractional subhalo
mass only, and are independent of the particular host or subhalo mass.

The radial distribution of satellite-satellite subhalo mergers closely follows
the radial distribution of subhaloes. This implies that most of the subhalo
mergers happen in the outer layers of the halo. For the most part, these
mergers involve subhaloes that are already dynamically associated before
accretion into the main halo, i.e.  they were either part of the same halo, or
of two separate haloes that were accreted at similar times and locations. At
every redshift, most of these subhaloes which subsequently merged were closer
together than $30\deg$ as seen from the centre of the halo that hosts the
merger, at the time they fell in.

Finally, we find that a small fraction of the high-mass subhaloes has
experienced a merger with another subhalo since accretion into the current host
halo. The values depend on the mass ratio of the merger, but vary from a few
percent for mass ratios greater than $0.3$ to $\sim 10\%$ for mass ratios greater
than $0.03$.

In spite of using some of the largest simulations to date, our results could
still be affected to some extent by numerical resolution. Due to the rarity of
the events we are trying to study, it is difficult to find a range of
substructure and host halo masses where we have, at the same time, (i) enough
particles to resolve substructures well, (ii) enough haloes to distinguish real
trends from cosmic variance, and (iii) enough subhaloes to establish their
properties and dynamics. Fortunately, as we have shown, many properties can be
described as a function of only the fractional subhalo mass. In these cases we
are observing the same system resolved with many different numbers of
particles, so it is reassuring that we find the same trends for different host
halo masses. This gives us confidence that these results are robust. On the
other hand, quantities which scale with halo mass are much less reliable and
could still be affected by resolution effects. Much larger simulations,
currently beyond reach, will be needed to check them.

\section*{Acknowledgements}
We are grateful to Adrian Jenkins and John Helly for providing us with
the high resolution simulation and merger trees used in this paper. We
also acknowledge Phil Bett, Liang Gao and Shaun Cole for helpful
discussions, and Lydia Heck for indispensable computing support.  The
Millennium simulation was carried out as part of the programme of the
Virgo Consortium on the Regatta supercomputer of the Computing Centre
of the Max-Planck Society in Garching. REA is supported by a
PPARC/British Petroleum sponsored Dorothy Hodgkin postgraduate
award. CMB is funded by a Royal Society University Research
Fellowship. CSF acknowledges a Royal Society-Wolfson Research Merit
award. This work was supported in part by a rolling grant from STFC to
the ICC.

\bibliographystyle{mn2e}
\bibliography{sub}

\section*{Appendix: numerical effects}

\begin{figure}
\begin{center}
\includegraphics[width=8.5cm]{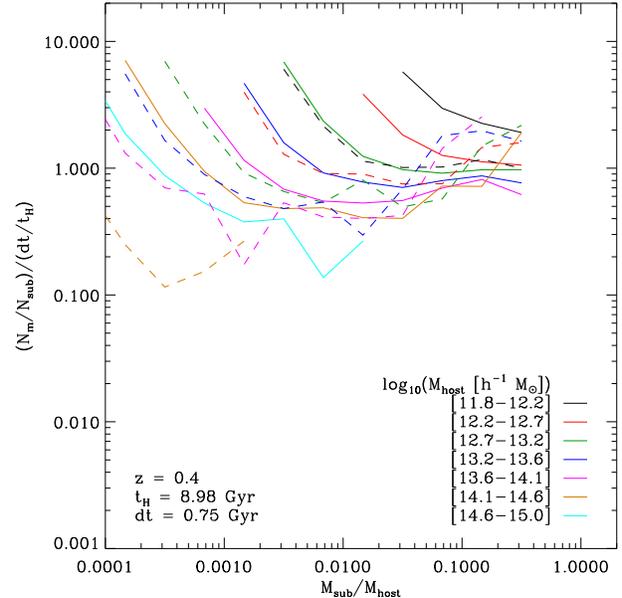}
\caption[The mean number of satellite-central subhalo mergers per
subhalo and per unit of time as a function of the subhalo mass.]{
The mean number of satellite-central subhalo mergers per
subhalo and per unit of time as a function of the subhalo mass. The
solid lines show the results from the MS while the dashed lines show
the result from the HS.  The coloured lines represent the results from
haloes of different mass, as indicated by the legend. Note that we
display results from subhaloes with 20 particles or more. The upturn in
$N_{m}$ for low mass subhaloes is due to the inclusion of subhaloes
resolved with fewer than 200 particles. Once the $N > 200$ criterion
is applied, the upturn disappears as shown in Fig \ref{sub:fig:subsub}.
\label{sub:fig:upturn}}
\end{center}
\end{figure}

Numerical artifacts can pose serious problems in obtaining a robust
estimate of various properties of the population of subhaloes. For
instance, two-body encounters, particle heating, or force softening
could easily dilute substructures that are not resolved with enough
particles \citep{Moore1996}. These problems translate into an
overestimation of the number of satellite-central subhalo mergers in
each timestep.

Such a feature is clear in Fig.~\ref{sub:fig:upturn}, which is similar to
Fig.~\ref{sub:fig:subsub}, but for satellite-central mergers only and
including subhaloes with less than $200$ particles. For these objects,
we can see a strong disagreement between the merger rate of
substructures in the simulations with different resolution which is
manifest as an upturn in the curves. However, the upturn disappears
for subhaloes with $N > 200$ which is the limit set in this paper.

An overestimation of the destruction rate also has implications for
other quantities such as the abundance and radial distribution of
subhaloes. For instance, the subhalo mass function shows a cut-off at
low masses compared with the expected power-law behaviour when we
include subhaloes resolved with fewer than $\sim 50$ particles. (This
quantity is less affected since most of the subhaloes are in the outer
layers of the halo.)  On the other hand, the inner part of the
radial distribution is more sensitive to these effects. Once subhaloes
with fewer than 200 particles are included in Fig.~\ref{sub:fig:rad_dist}
the distribution becomes less centrally concentrated.

Our convergence study indicates that 200 particles is the limit below
which results are unduly affected by resolution. This is why we have
adopted this minimum particle count throughout this chapter, except when
otherwise stated explicitly. This choice should minimize
finite-resolution effects.

\label{lastpage}
\end{document}